\documentclass[aps,prd,showpacs,amssymb,twocolumn,nofootinbib]{revtex4}
\usepackage{graphicx}
\newcommand{\bp}{\bar{\psi}}
\newcommand{\bmp}{p_{\bar{\psi}}}
\newcommand{\bpp}{\bar{P}_{\bp}}

\begin{document}
\title{Particles and vacuum for perturbative and
non-perturbative Einstein-Rosen gravity}

\author{J. Fernando \surname{Barbero G.}}
\affiliation{Instituto de Matem\'{a}ticas y F\'{\i}sica Fundamental, CSIC,
Serrano 113bis, 28006 Madrid, Spain}
\author{Guillermo A. \surname{Mena Marug\'an}}
\affiliation{Instituto de Estructura de la Materia, CSIC, Serrano
121, 28006 Madrid, Spain}
\author{Eduardo J. \surname{S. Villase\~nor}}
\affiliation{Dpto. Matem\'{a}ticas, Escuela Polit\'{e}cnica Superior,
Universidad Carlos III de Madrid, Avda. de la Universidad 30,
28911 Legan\'{e}s, Spain}

%\date{May 26, 2004}

\begin{abstract}

We discuss the connection between the Fock space introduced by
Ashtekar and Pierri for Einstein-Rosen waves and its perturbative
counterpart based on the concept of particle that arises in
linearized gravity with a de Donder gauge. We show that the gauge
adopted by Ashtekar and Pierri is indeed a generalization of the
de Donder gauge to full (i.e. non-linearized) cylindrical gravity.
This fact allows us to relate the two descriptions of the
Einstein-Rosen waves analyzed here (the perturbative one and that
made by Ashtekar and Pierri) by means of a simple field
redefinition. Employing this redefinition, we find the highly
non-linear relation that exists between the annihilation and
creation-like variables of the two approaches. We next represent
the particle-like variables of the perturbative approach as
regularized operators, introducing a cut-off. These can be
expanded in powers of the annihilation and creation operators of
the Ashtekar and Pierri quantization, each additional power being
multiplied by an extra square-root of ($\hbar$ times) the
three-dimensional gravitational constant, $\sqrt{G}$. In
principle, the perturbative vacuum may be reached as the limit of
a state annihilated by these regularized operators when the
cut-off is removed. This state can be written as the vacuum of the
Ashtekar and Pierri quantization corrected by a perturbative
series in $\sqrt{G}$ with no contributions from particles with
energies above the cut-off. We show that the first-order
correction is in fact a state of infinite norm. This result is
interpreted as indicating that the Fock quantizations in the two
approaches are unitarily inequivalent and, in any case, proves
that the perturbative vacuum is not analytic in the interaction
constant. Therefore a standard perturbative quantum analysis
fails.

\end{abstract}


\maketitle
\renewcommand{\thesection}{\Roman{section}}
\renewcommand{\theequation}{\arabic{section}.\arabic{equation}}

\section{Introduction}

Solutions to general relativity in vacuo with whole cylindrical
symmetry seem to have been first found by Beck \cite{Be} and then
rediscovered by Einstein and Rosen (ER) in the search for
spacetimes that could describe the propagation of gravitational
waves \cite{ER}. By whole cylindrical symmetry \cite{WCS} we
understand the existence (in topologically trivial spacetimes) of
two linearly independent, commuting, and hypersurface orthogonal
Killing vector fields, one of them rotational and the other
translational. Among the motivations for the study of these
solutions was Einstein's belief that one of the fundamental
problems of physics (at least at his time) was the lack of a
satisfactory theory of radiation, specially in the presence of the
gravitational field \cite{Ch}.

The ER solutions are cylindrical gravitational waves with linear
polarization. Cylindrical waves with general polarization, whose
Killing vector fields are not hypersurface orthogonal, were
originally analyzed by Ehlers and collaborators and by Kompaneets
\cite{EK}.

The ability to provide a model with the field complexity of
general relativity, but with known exact solutions which describe
gravitational waves, has endowed the family of ER spacetimes with
a prominent role in the analysis of the quantization of
gravitational systems
\cite{Ku,Al,AP,Ma,AMM,Ash,GP,DT,BMV,Ni2,KHB}. Kucha\v{r} pioneered
this line of work by discussing the canonical quantization of
these cylindrical waves \cite{Ku}. A key remark in this discussion
is that the dynamics of the ER spacetimes is equivalent to that of
a cylindrically symmetric, massless scalar field propagating on an
auxiliary Minkowski background. Thanks to this fact, one can
recast the system as three-dimensional gravity coupled to a scalar
field with rotational symmetry. This was precisely the approach
followed by Allen to further explore the quantization of the
model, studying regularization issues and the relevance of the
quantum fluctuations around the vacuum \cite{Al}.

Employing this three-dimensional formulation, a consistent and
essentially complete quantization of the ER waves was obtained
some years ago by Ashtekar and Pierri (AP) \cite{AP}. This
quantization was achieved after a careful treatment of the
regularity conditions at the symmetry axis, on the one hand, and
of the boundary conditions at spatial infinity that ensure
asymptotic flatness in cylindrical gravity \cite{AP,RT}, on the
other. The quantization accounts as well for certain functional
analytic subtleties that arise in the regularization of metric
operators. Actually, some of these subtleties were later revisited
by Varadarajan \cite{Ma}. The definition and regularization of the
metric operators, not from the perspective of three-dimensional
gravity coupled to a scalar field, but from a purely
gravitational, four-dimensional perspective was discussed in Ref.
\cite{AMM}.

This quantum framework has allowed to show that there exist
unexpectedly large quantum effects in the model, at least in the
asymptotic region \cite{Ash,GP,DT} and the cylindrical axis
\cite{AMM}. A more detailed study of the consequences of the
vacuum fluctuations for microcausality, including the smearing of
light cones all over the spacetime and the blurring of the
symmetry axis, has recently been carried out in Ref. \cite{BMV}.

In the quantization proposed by Ashtekar and Pierri, the Hilbert
space is the Fock space corresponding to the rotationally
symmetric scalar field that propagates in the three-dimensional,
auxiliary Minkowski spacetime. There exist two relevant notions of
evolution in this Hilbert space: one associated with the auxiliary
Minkowski time and another with the physical time \cite{AP,BMV}.
In the former case, the dynamics is dictated by the Hamiltonian of
the axisymmetric, massless scalar field, $H_0$, while in the
latter the Hamiltonian is a non-linear, bounded function of it,
$H=(1-e^{-4G_3H_0})/(4G_3)$ \cite{AV,Va,ABS}. Here, $G_3$ is the
three-dimensional gravitational constant or, equivalently, the
effective Newton constant per unit length in the direction of the
axis \cite{AMM}.

The difference between the dynamical generators arises because the
presence of energy in the gravitational waves causes a deficit
angle at spatial infinity that affects the norm of the asymptotic,
timelike Killing vector. Since this norm must be unity for the
physical time, one must consider an energy dependent change of
time that leads to the above transformation in the Hamiltonian. In
fact, the emergence of a bounded physical Hamiltonian proportional
to the deficit angle produced by the wave is a feature not just of
ER gravity, but of cylindrical gravitational waves with general
polarization (even in the presence of spinning strings)
\cite{Gu,MMM}.

Regardless of which Hamiltonian is considered to govern the
quantum dynamics, $H_0$ or $H$, the Fock spaces and quantizations
obtained in both cases are equivalent, inasmuch as the
corresponding evolution operators are unitary and the two initial
time copies are exactly the same \cite{BMV}. However, a question
that has not been addressed yet in the literature is whether the
Fock space employed in the AP quantization is the kind of Fock
space that one would introduce in a perturbative treatment of the
ER model and, if they differ, what relation exists between them.
The main aim of this article is to discuss this issue. This is a
fundamental question in order to answer whether one can attain or
not the correct non-perturbative results by adopting a
perturbative approach.

In a perturbative formalism, one would adopt as metric variables
linear combinations of the difference between the Minkowski
background and the actual spacetime metric, expanding the
gravitational action in powers of them. The quadratic term
provides the action of linearized gravity, while the higher-order
terms can be regarded as describing interactions. At this stage,
it is convenient to adopt a gauge that simplifies the linearized
equations. A frequently used gauge is the de Donder or Lorentz
gauge \cite{MTW}, in which the linearized gravitational equations
reduce to wave equations, so that one easily arrives at a notion
of particle.

We will see that the gauge fixing introduced by Ashtekar and
Pierri is nothing but a generalization of the de Donder gauge from
linearized to full ER gravity. Therefore, adopting it as a valid
gauge (with a clear interpretation in linearized gravity), the
relation between the AP and the perturbative treatments will
straightforwardly follow from the transformation on the
configuration space that connects the metric variables used in
each of the two descriptions. In particular, this transformation,
when completed into a canonical one, will provide the relation
between the particle-like variables of the two formalisms.

The plan of the work is as follows. We first review the ER model
and the most important aspects of the AP quantization in Sec. II.
In Sec. III we introduce a description of the ER waves in terms of
fields that are linear in the excess of the metric around
Minkowski and translate to them the AP gauge fixing. In Sec. IV we
adapt to this description the discussion of Ref. \cite{BMV} about
the linearization of the model. Sec. V proves that the
linearization of the AP gauge is a de Donder gauge. Furthermore,
while in general relativity the de Donder gauge leaves some
ambiguity in the choice of coordinates \cite{MTW}, the gauge is
completely determined in linearized ER gravity when one imposes
suitable regularity conditions, corresponding to a fixed location
of the symmetry axis. Employing the transformation that maps the
basic metric field of the reduced ER model in our description
(linear in the excess around Minkowski) to the axisymmetric scalar
field of the AP approach, we find in Sec. VI the relation between
the creation and annihilation variables associated with each of
the two fields considered. These variables are promoted to
regularized operators in Sec. VII. Using them, we try to determine
the perturbative vacuum in Sec. VIII. In particular, we
investigate whether this vacuum can be reached from the
non-perturbative one by means of a series expansion in (the
square-root of) the gravitational constant. The answer turns out
to be in the negative, because the first correction to the
non-perturbative vacuum would have an infinite norm, even if
ultraviolet divergences are regulated with a cut-off. Finally,
Sec. IX contains the conclusions.

In the rest of the paper, we call $G=G_3\hbar$, and adopt a system
of units in which $\hbar=c=1$, with $\hbar$ being Planck constant
and $c$ the speed of light. Note that $G_3$ is an inverse energy,
whereas $G$ has dimensions of length.

\section{The Ashtekar-Pierri quantization}
\setcounter{equation}{0}

The ER waves are linearly polarized, cylindrical waves in vacuum
general relativity. They can be described by the metric
\cite{AP,AMM}:
\begin{eqnarray}\label{met}
ds^2&=&e^{-\psi}\left[-N^2
dt^2+e^{\gamma}(dR+N^{R}dt)^2+(8Gr)^2d\theta^2\right]\nonumber\\
&+&e^{\psi}dZ^2.\end{eqnarray} Here, $Z\in \mathbb{R}$ is the
coordinate of the symmetry axis, $\theta\in S^1$ corresponds to
the axial coordinate, $R\in \mathbb{R}^+$ will be called the
radial coordinate, $N^R$ describes the radial component of the
shift vector and $N$ is the lapse function. Owing to the
cylindrical symmetry, all metric functions $N^R$, $N$, $\psi$,
$r$, and $\gamma$ depend only on the time $t$ and on the radial
coordinate $R$. We follow the convention that the dimensionality
of the spacetime interval is carried by the coordinates $t$, $R$,
and $Z$, while the metric fields are dimensionless \cite{Di}.

The gravitational action, endowed with surface terms suitable for
a fixed metric on the boundaries (namely, the initial and final
$t$-sections and an exterior cylinder $R=R_f$ in the limit
$R_f\rightarrow\infty$ \cite{AMM}), takes the form \cite{Ku,RT}
\begin{equation}\label{act}
S=\int_{t_1}^{t_2} dt \left[-{\cal H}+\!\int_0^\infty dR\,
(p_{\gamma}\dot{\gamma\;}+p_r\dot{r}+p_{\psi}\dot{\psi})
\right],\end{equation} where the $p$'s are the momenta canonically
conjugate to the metric fields, the dot stands for the derivative
with respect to $t$, and $\cal{H}$ is the total Hamiltonian
\[
{\cal H}=\frac{1}{4G_3}\left(1-e^{-\gamma_{\infty}/2}\right)\!+\!
\int_0^\infty \!\!\!dR\left[NC+N^RC^R\right]\!.\] The first
contribution is a boundary term at infinity, with
$\gamma_{\infty}:= \gamma(R\rightarrow\infty)$. Besides, the
Hamiltonian constraint $C$ and the (radial) diffeomorphism
constraint $C^R$ are:
\begin{eqnarray}
\hspace*{-.4cm}C\!&=&\!\!e^{-\gamma/2}\!\left[2r^{\,\prime\prime}
-\gamma^{\prime}r^{\:\prime}\!-p_{\gamma}p_{r}+\frac{p_{\psi}^2}
{2r}+\frac{r(\psi^{\prime})^2}{2}\right]\!,\nonumber\\
\hspace*{-.4cm}C^R\!&=&\!-2p^{\prime}_{\gamma}+
p_{\gamma}\gamma^{\prime}+p_rr^{\:\prime}+p_{\psi}\psi^{\prime}
.\nonumber\end{eqnarray} The prime denotes the derivative with
respect to $R$.

The Lagrangian of the model can be straightforwardly obtained by
means of a Legendre transformation using the relation between
momenta and time derivatives of the metric that the Hamilton
equations provide:
\begin{eqnarray}\label{haeq}
p_{\gamma}Ne^{-\gamma/2} &=& -\dot{r}+N^Rr^{\:\prime}, \nonumber\\
p_r N e^{-\gamma/2}&=&-\dot{\gamma\,}+2\left(N^R\right)
^{\prime}-N^R\gamma^{\prime},\nonumber\\ p_{\psi} Ne^{-\gamma/2}
&=& r\,\dot{\psi}- N^Rr\,\psi^{\prime}.\end{eqnarray}

All metric fields, as well as their momenta, are subject to
boundary conditions that ensure the regularity at the symmetry
axis, the asymptotic flatness at spatial infinity (with a possible
deficit angle), and a well-posed Hamiltonian dynamics. We refer to
the work of Ashtekar and Pierri \cite{AP} for details about these
conditions. In particular, we asume that all fields are
$C^{\infty}$ everywhere, that at the axis
\begin{eqnarray}\label{regu}
\gamma(R=0)&=&0,\hspace*{.4cm}N^R(R=0)=0,\nonumber\\
\rho(R=0)&=&0,\hspace*{.4cm} \rho^{\prime}(R=0)=0, \end{eqnarray}
where we have called $\rho:=r-R/(8G)$, and that at spatial
infinity
\begin{equation}\label{asym}
\psi=O\left(\frac{1}{R}\right),\hspace*{.4cm}
\rho=O(1).\end{equation} We say that a function $f$ is of
asymptotic order $O(R^{-a})$ if the products $R^af$,
$R^{a+1}f^{\prime}$, and $R^{a+2}f^{\prime\prime}$ admit limits as
$R$ tends to infinity \cite{AP}. In addition, we note that, for
the stability under diffeomorphisms of the regularity condition
that $\gamma$ vanish at the axis, one must further restrict the
shift vector to satisfy \cite{Ma2}
\begin{equation}\label{regu2}
\{\gamma(\tilde{R}),\int dR \,
N^RC^R\}|_{\tilde{R}=0}=2\big(N^R\big)^{\prime}|_{\tilde{R}=0}=0,
\end{equation}
where the curved brackets denote Poisson brackets.

As shown by Ashtekar and Pierri \cite{AP}, the gauge freedom
corresponding to the Hamiltonian and radial constraints can be
totally removed by imposing, respectively, the conditions
\begin{equation}\label{conAP}
\chi^R:=r-\frac{R}{8G}=0,\hspace{.6cm}\chi\,:=p_{\gamma}=0.
\end{equation}
In this way, one arrives at a reduced model whose only degree of
freedom is the scalar field $\psi$. The reduced metric is given by
Eq. (\ref{met}) with $N^R=0$, $8Gr=R$ and
\begin{equation} \label{gamma}
\gamma= \int_0^R d\tilde{R}\,\frac{\tilde{R}}{2}
\left[(\psi^{\prime})^2+\frac{(8Gp_{\psi})^2}{\tilde{R}^2}
\right].\end{equation} The reduced dynamics in the time $t$ is
generated by the physical Hamiltonian $H=(1-e^{-4G_3H_0})/(4G_3)$,
where $H_0:=\gamma_{\infty}/(8G_3)$ is (up to a constant factor)
the Hamiltonian that would correspond to a free, axisymmetric,
massless scalar field $\psi$ in three dimensions.

The evolution of the field $\psi$ gets considerably simplified if
one introduces the energy dependent change of time
$T=e^{-4G_3H_0}t$. In this auxiliary time, the dynamics is
dictated precisely by the free-field Hamiltonian $H_0$; so the
equation of motion for $\psi$ is just a wave equation with
rotational symmetry in a three-dimensional Minkowski spacetime
with coordinates $(T,R,\theta)$. The classical solutions with
regularity at the axis have the form
\[
\psi(R,T)\!=\!\sqrt{4G}\!\int_0^\infty \!\!\!dk\, J_0(Rk)\!
\left[\!A(k)e^{-ikT}\!\!+\!A^\dagger(k)e^{ikT}\right]\!,
\]
where $J_0$ is the zeroth-order Bessel function \cite{AG}. The
function $A(k)$ and its complex conjugate $A^\dagger(k)$ are fixed
by the initial conditions and play the role of annihilation and
creation variables. In terms of them, the free-field Hamiltonian
can be written as $H_0=\int_0^{\infty}dk\,kA^{\dagger}(k)A(k)$.

The quantization of the reduced ER model can be carried out
following standard techniques. We introduce a Fock space in which
$\hat{\psi}(R,0)$, the quantum counterpart of $\psi(R,0)$, is an
operator-valued distribution \cite{RS}. Its action is determined
by those of $\hat{A}(k)$ and $\hat{A}^\dagger(k)$, annihilation
and creation operators with non-vanishing commutators
\begin{equation}\label{ancre}
\left[\hat{A}(k),\hat{A}^\dagger(\tilde{k})\right]=
\delta(k,\tilde{k}).
\end{equation}
Explicitly,
\begin{equation}\label{distri}
\hat{\psi}(R,0)\!=\!\sqrt{4G}\int_0^\infty \!\!dk\,
J_0(Rk)\!\left[\hat{A}(k)\!+\!\hat{A}^\dagger(k)\right]
\end{equation}
and the Fock space is constructed over the Hilbert space of square
integrable functions on the positive real axis,
$L^2(\mathbb{R}^+,dk)$. In this space, the free-field Hamiltonian
is represented by the self-adjoint operator
\[
\hat{H}_0=\int^{\infty}_0dk\,k\,\hat{A}^{\dagger}(k)\hat{A}(k).
\]
Via the spectral theorem, we can then promote the physical
Hamiltonian to a bounded operator $\hat{H}$. The evolution in the
time parameters $T$ and $t$ are thus respectively given by the
unitary operators $\hat{U}_0(T)=e^{-iT\hat{H}_0}$ and
$\hat{U}(t)=e^{-it\hat{H}}$.

In order to arrive at a well-defined operator for $\psi$, it is
necessary to regularize the quantum field (\ref{distri})
\cite{AMM}. This can be done by smearing the Bessel function
$J_0(Rk)$ with a square integrable function $g\in
L^2(\mathbb{R}^+,dk)$,
\begin{equation}\label{regfi}
\hat{\psi}(R|g)\!=\!\sqrt{4G}\int_0^\infty \!\!dk\,g(k)
J_0(Rk)\!\left[\hat{A}(k)\!+\!\hat{A}^\dagger(k)\right]\!.
\end{equation}
It is then possible to define meaningful metric operators, e.g.
for the diagonal $Z$ and $\theta$ components (at the initial
time), by exponentiating the regulated version of the field,
$e^{\pm\hat{\psi}(R|g)}$ \cite{AMM}. For simplicity, we will
restrict in the following to regulators that correspond to a
cut-off $k_c$, so that
\begin{equation}\label{cut}
g(k)=\left\{\begin{array}{ll}1 & {\rm if}\; k\leq k_c\,,\\ 0 &
{\rm if}\; k> k_c\,.\end{array}\right.\end{equation} In
particular, we have $g^2(k)=g(k)$.

\section{New metric fields}
\setcounter{equation}{0}

We will introduce now a different field parametrization for the
Einstein-Rosen metric such that the new fields are linear in the
metric excess around Minkowski,
\begin{eqnarray}\label{neme}
ds^2\!&=&\!\!-(1-2\bar{N}-\bar{\psi})dt^2+2\bar{N}^RdtdR+
(1+\bar{\gamma}-\bar{\psi})dR^2\nonumber\\&+&(R^2-R^2\bar{\psi}
+16GR\bar{\rho})d\theta^2+(1+\bar{\psi})dZ^2.\end{eqnarray}
Obviously, when all the fields
$(\bar{N},\bar{N}^R,\bar{\psi},\bar{\rho},\bar{\gamma})$ vanish,
we recover the Minkowski solution. Moreover, our new
parametrization is specially suited for the study of linearized
gravity, since it reproduces the linearization of the spacetime
metric (\ref{met}) around Minkowski \cite{BMV} [apart from the
notation with over-bars and with $\rho=r-R/(8G)$]. Hence, at first
order one can interpret our new metric fields as the linearization
of those used in the AP formulation.

The exact relation between both sets of fields is found by
identifying the metric expressions. One gets
\begin{eqnarray}\label{neps}\psi\!&=&\!\ln{(1+\bar{\psi})},\\
\gamma\!&=&\!\ln{\left[(1+\bar{\psi})(1+\bar{\gamma}-
\bar{\psi})\right]},\nonumber\\
8Gr\!&=&\!\sqrt{1+\bar{\psi}}\,\sqrt{R^2-R^2\bar{\psi}+
16GR\bar{\rho}}\,,\nonumber\\ N\!&=&\!\sqrt{1+\bar{\psi}}\,
\sqrt{1+2\bar{N}-\bar{\psi}-\frac{(\bar{N}^R)^2}
{1+\bar{\gamma}-\bar{\psi}}}
\,,\nonumber\\N^R\!\!\!&=&\!\frac{\bar{N}^R}{1+\bar{\gamma}-\bp}.
\label{nefi}
\end{eqnarray}
In principle, if one insists on imposing that the fields on the
left hand-side be real, the range of the new metric fields should
be properly restricted. Note also that the transformation is
always well-defined (even with these reality conditions) in a
neighborhood of the Minkowski solution, i.e. for small over-barred
fields and hence for the linearized theory.

The above transformation can be easily completed into a canonical
one. Old and new momenta are related by
\begin{eqnarray}\label{nemo}
p_{\psi}\!&=&\!(1+\bp)\,\bmp+(2\bp-\bar{\gamma})
\,p_{\bar{\gamma}}
+\left(\frac{R}{8G}\bp-\bar{\rho}\right)p_{\bar{\rho}},
\nonumber\\
p_{\gamma}\!&=&\!
(1+\bar{\gamma}-\bp)\,p_{\bar{\gamma}},\nonumber\\
p_{r}\!&=&\!\sqrt{1-\bp+\frac{16G\bar{\rho}}{R}}\;
\frac{p_{\bar{\rho}}}
{\sqrt{1+\bp}}.\end{eqnarray}

Using these expressions, it is not difficult to write the
constraints of the ER model in terms of the new canonical
variables. The associated gauge freedom can be eliminated by
imposing conditions equivalent to those employed by Ashtekar and
Pierri, namely
\begin{equation}\label{negafi}
\bar{\chi}^R:=(1+\bp)\bar{\rho}-\frac{R}{16G}\bp^2=0,
\hspace*{.4cm}\bar{\chi}:=
p_{\bar{\gamma}}=0.\end{equation} Taking into account relations
(\ref{neps}), (\ref{nefi}), and (\ref{nemo}), one can check that
these conditions are identical to the gauge fixing requirements
(\ref{conAP}) except for metric-dependent, global multiplicative
factors that differ from zero for almost all values of the metric
fields and, in particular, in a neighborhood of their origin (the
Minkowski background).

The reduction of the model can be carried out by the same
procedure followed by Ashtekar and Pierri, translated to the new
variables, so we will not repeat the details here. Perhaps the
only noticeable point is that, owing to the dependence of the
gauge fixing condition $\bar{\chi}^R$ on $\bp$, the momentum
canonically conjugate to $\bp$ after reduction, $\bar{P}_{\bp}$,
does not coincide with the original one $\bmp$ (in other words,
the Poisson and Dirac brackets of $\bp$ and $\bmp$ differ). One
finds instead
\[
\bar{P}_{\bp}=\frac{2(1+\bp)}{2(1+\bp)+R\,
\bp\bp^{\prime}(2+\bp)^{^{\;}}}
\,p_{\bp}.
\]

The only physical degree of freedom is described by $\bp$. The
reduced metric, regular at the axis and with asymptotically unit
lapse, takes the form
\begin{eqnarray}
ds^2\!&=&\!(1+\bar{\gamma}-\bp)\left[-\frac{dt^2}{1+
\bar{\gamma}_{\infty}}+dR^2\right]+\frac{R^2}{1+\bp}
\,d\theta^2\nonumber\\
&+&(1+\bp)dZ^2,\nonumber\end{eqnarray} with $\bar{\gamma}$ given
by
\begin{eqnarray}
\hspace*{-.5cm}&&\hspace*{-.5cm}(1+\bar{\gamma}-\bp)(1+\bp)=
\nonumber\\
\hspace*{-.5cm}&&\hspace*{-.3cm}\exp{\!\left\{\!\int_0^{R}
\!\!\!\!d\tilde{R}\,
\frac{\tilde{R}}{2}\!\left[\!\left(\frac{\bp^{\prime}}
{1+\bp}\right)^2\!\!+\!\left(\frac{8G\bar{P}_{\bp}}
{\tilde{R}}\right)^2\!\!(1+\bp)^2\right]\!\!\right\}}.
\label{baga}
\end{eqnarray}
In the reduced metric, $\bar{\gamma}_{\infty}$ denotes the limit
when $R\rightarrow\infty$ of $\bar{\gamma}$. Recalling conditions
(\ref{asym}) and relations (\ref{nefi}), which in particular imply
that $\bp_{\infty}=0$, one easily checks that
\begin{eqnarray}
\hspace*{-.5cm}&&\hspace*{-.4cm}
1+\bar{\gamma}_{\infty}=e^{\gamma_{\infty}}=\nonumber\\
\hspace*{-.5cm}&&\hspace*{-.3cm}\exp{\!\left\{\!\int_0^{\infty}
\!\!\!\!d R\,
\frac{R}{2}\left[\!\left(\frac{\bp^{\prime}}{1+\bp}\right)^2
\!\!\!+\!\left(\frac{8G\bar{P}_{\bp}}{R}\right)^2
\!\!(1+\bp)^2\right]\!\!\right\}}.
\label{bagain}
\end{eqnarray}

Finally, the action of the reduced model is
\[
S_r\!=\!\!\int_{t_1}^{t_2}\! dt\!
\left[-\frac{1}{4G_3}\left(\!1-\!
\frac{1}{\sqrt{1+\bar{\gamma}_{\infty}}}\!\right)\!+
\!\int_{0}^{\infty}\!dR\, \bar{P}_{\bp}\dot{\bp\;}\right]\!.\] The
first contribution is therefore minus the reduced Hamiltonian. In
fact, as one could expect, the above expressions for the reduced
metric and action reproduce exactly those obtained in the AP
treatment of the ER model, provided that the basic fields $\psi$
and $\bp$ are related by Eq. (\ref{neps}) [$\psi=\ln{(1+\bp)}$]
and their canonical momenta by
\begin{equation}\label{motra}
p_{\psi}=\bar{P}_{\bp}(1+\bp).\end{equation} This is precisely the
change of momentum needed to complete our field redefinition into
a canonical transformation for the reduced system. In other words,
instead of changing the metric fields for the ER waves, completing
the transformation into a canonical one, and reducing the system,
one can simply proceed to redefine the field $\psi$ in the reduced
model while preserving its canonical symplectic structure.

It is worth noting that, if one insists that the induced metric of
the reduced model be positive, one should demand that the field
$\psi$ be real or, equivalently, that $\bp>-1$. In a perturbative
analysis, however, the field $\bp$ is directly assumed to be small
and so the above reality condition would be obviated in practice.

\section{Einstein-Rosen waves in linearized gravity}
\setcounter{equation}{0}

Let us consider the linearization of the ER model around the
Minkowski solution. Since our new parame\-trization of the metric
is linear in the excess around the flat background, the action of
the linearized theory $\bar{S}_l$ can be obtained from that in
general relativity by keeping only up to quadratic terms in our
fields. Our discussion will essentially follow the lines presented
in Ref. \cite{BMV}. One starts with the Lagrangian form of the
action (\ref{act}), that can be easily deduced employing the
Hamilton equations (\ref{haeq}). Next, one expresses it in terms
of our new metric fields employing the transformations
(\ref{neps}) and (\ref{nefi}), and expands the result in powers of
those fields. As we have commented, the action $\bar{S}_l$ is
given by the terms quadratic in the fields. In deducing this
action, one can get rid of several boundary terms by using the
regularity and asymptotic conditions (\ref{regu}) and
(\ref{asym}). Together with our field redefinitions, the latter of
these sets of conditions implies that, at spatial infinity, the
limits of $\bp$, $R\bp^{\prime}$, $\bar{\rho}/R$, and
$\bar{\rho}^{\prime}$ vanish. The regularity conditions
(\ref{regu}) guarantee in turn that [for generic values of
$\bp(R=0)$] $\bar{N}^R$ and $\bar{\rho}$ vanish at the axis and
\begin{equation}\label{lic}
\bar{\gamma}(R\!=\!0)\!=\!16G\bar{\rho}^{\,\prime}(R\!=\!0)\!=\!
\frac{\bp^2(R\!=\!0)} {1+\bp(R\!=\!0)}.\end{equation} We notice
also that the strict linearization of these relations between
field values at the symmetry axis leads to
$\bar{\gamma}(R\!=\!0)\!=\!\bar{\rho}^{\,\prime}(R\!=\!0)\!=0$.

Writing back the action in Hamiltonian form by means of a Legendre
transformation, one gets \cite{BMV}
\begin{eqnarray}\label{liact}
\bar{S}_l\!&=&\!\!\!\int_{t_1}^{t_2} dt\left[-\bar{\cal H}_l +\!
\int_0^\infty dR\, (P_{\bar{\gamma}}
\dot{\bar{\gamma}\,}+P_{\bar{\rho}}\dot{\bar{\rho}\,}+
P_{\bp}\dot{\bp\,})
\right]\!,\\ \bar{\cal H}_l\!&=&\!\!\!\int_0^\infty\!\! dR
\!\left[
\frac{4GP_{\bp}^2}{R}\!+\frac{R\,(\bp^{\prime})^2}{16G}-
\!P_{\bar{\gamma}}
P_{\bar{\rho}}+\bar{N}\bar{C}_l+\!\bar{N}^R\bar{C}^R_l\right]\!.
\nonumber
\end{eqnarray}
Here, the $P$'s denote the canonical momenta in the linearized
model and, in terms of time derivatives of the metric, take the
expressions
\begin{equation}\label{lihameq}
P_{\bar{\gamma}}\!=\!\frac{\bar{N}^R}{8G}
-\dot{\bar{\rho}\,},\hspace*{.4cm}P_{\bar{\rho}} \!=
\!2\,(\bar{N}^R)^{\prime}-\dot{\bar{\gamma}\,},\hspace*{.4cm}
P_{\bp}\!=\!\frac{R}{8G}\dot{\bp\;}.\end{equation} In addition,
the linearized Hamiltonian and radial constraints are:
\begin{eqnarray}\label{licon}
\bar{C}_l=2\bar{\rho}^{\,\prime\prime}-
\frac{\bar{\gamma}^{\prime}}{8G},
\hspace{.6cm}\bar{C}^R_l=
\frac{P_{\bar{\rho}}}{8G}-2P_{\bar{\gamma}}^{\prime}.\end{eqnarray}

It has recently been shown \cite{BMV} that the gauge freedom of
the linearized system can be completely removed by demanding that
\[
\bar{\chi}^R_l:=\bar{\rho}=0,\hspace*{.4cm}
\bar{\chi}_l:=P_{\bar{\gamma}}=0.\] We point out that these
conditions are just the linearization of the gauge fixing
requirements (\ref{negafi}) imposed on the full ER model or,
equivalently, of the gauge fixing (\ref{conAP}) introduced by
Ashtekar and Pierri. The subsequent reduction can be carried out
exactly as explained in Ref. \cite{BMV}. The degrees of freedom of
the system are $\bp$ and its momentum. The reduced action is
\begin{eqnarray*}
\bar{S}_2=\int_{t_1}^{t_2} dt\left[- H_0+ \int_0^\infty\! dR\;
P_{\bp}\dot{\bp\;}\right].\end{eqnarray*} Here, $H_0$ is again the
Hamiltonian of a massless scalar field with rotational symmetry in
three dimensions, and can be identified with
$\gamma_{\infty}/(8G_3)$ if, in Eq. (\ref{gamma}), one evaluates
$\gamma$ at the canonical pair $(\bp,P_{\bp})$. At quadratic order
in these fields, we see from Eq. (\ref{bagain}) that $H_0$
coincides as well with $\bar{\gamma}_{\infty}/(8G_3)$. Thus, the
free-field Hamiltonian provides the reduced Hamiltonian of the
model in linearized gravity.

Moreover, in the selected gauge, the three-dimensional metric of
the linearized model is just Minkowski. The scalar field $\bp$
determines the norm of the Killing vector $\partial_{Z}$, and
appears in the reduced four-dimensional metric of the linearized
theory in the form
\begin{equation}
\label{mink} ds_l^2\!=\!(1-\bp)
(-dT^2+dR^2+R^2d\theta^2)+(1+\bp)dZ^2.\end{equation} We have
renamed $T$ the time coordinate of the linearized system to
emphasize that it can be identified with the auxiliary time of the
reduced ER model analyzed in Sec. II, inasmuch as they are both
Minkowskian in three dimensions and the corresponding evolution is
generated by the free-field Hamiltonian $H_0$ in both cases.

\section{De Donder gauge}
\setcounter{equation}{0}

We want to prove now that the gauge chosen by Ashtekar and Pierri
is a higher-order generalization of the de Donder gauge for
linearized gravity, i.e. that the linearization of the AP gauge
(which is precisely that imposed in Sec. IV to reduce the
linearized system) is of de Donder type. Let us start with some
notation. We call $h_{\mu\nu}=g_{\mu\nu}-\eta_{\mu\nu}$ the
difference between the actual spacetime metric $g_{\mu\nu}$ and
the flat metric $\eta_{\mu\nu}$ of the Minkowski background
[$\eta_{\mu\nu}=(-1,1,1,1)$ in a Cartesian coordinate system].
Standard perturbative approaches to gravity analyze the
gravitational interaction by means of an expansion in
\[
\bar{h}_{\mu\nu}=h_{\mu\nu}-\frac{h}{2}\eta_{\mu\nu}
\]
where $h=h_{\mu\nu}\eta^{\mu\nu}$ is the trace of $h_{\mu\nu}$. In
a first-order approximation (namely, in linearized gravity), only
terms up to quadratic in $\bar{h}_{\mu\nu}$ are maintained in the
action. To go beyond that approximation, higher-order terms are
treated perturbatively as interactions.

In the linearized theory, a frequently followed approach is to
(partially) remove the gauge freedom by introducing a set of
conditions that simplify the equations of motion for
$\bar{h}_{\mu\nu}$, transforming them into wave equations
\cite{MTW}, namely
\[
\bar{h}_{\mu\nu,\delta}\,\eta^{\nu\delta}=\bar{h}_{\mu\nu,}
^{\;\;\;\;\;\nu}=0.
\]
These conditions are known as de Donder (or Lorentz) gauge, and
provide an acceptable gauge fixing in linearized gravity, although
they still leave some freedom in the choice of coordinates
\cite{MTW}.

The most straightforward way to elucidate whether the gauge
imposed in the AP analysis of the ER waves is a higher-order
generalization of a de Donder gauge is to calculate the value of
$\bar{h}_{\mu\nu,}^{\;\;\;\;\;\nu}$ in the reduced model of Sec.
III. If $\bar{h}_{\mu\nu,}^{\;\;\;\;\;\nu}$ vanishes at
first-order in the fields, then the linearization of our gauge
fixing conditions is indeed a de Donder gauge.

Changing from cylindrical coordinates $(R,\theta)$ to Cartesian
ones $(x,y)$ and employing that, for any cylindrically symmetric
field $f$,
\begin{eqnarray*}
\partial_xf=\frac{x}{R}f^{\prime},\hspace*{.4cm} \partial_y
f=\frac{y}{R} f^{\prime},\hspace*{.4cm}
\partial_Z f=0,\end{eqnarray*}
a direct calculation shows that
$\bar{h}_{Z\nu,}^{\;\;\;\;\;\nu}=0$ and
\begin{eqnarray}\label{deDopr}
\bar{h}_{t\nu,}^{\;\;\;\;\;\nu}&=&-\frac{1}{2}\left[
\frac{\bar{\gamma}_{\infty}}
{1+\bar{\gamma}_{\infty}}(\dot{\bar{\gamma}\;}-\dot{\bp\;}\!)
+\frac{\dot{\bp\;}\bp\,(2+\bp)}{(1+\bp)^2}\right],\nonumber\\
\bar{h}_{x\nu,}^{\;\;\;\;\;\nu}&=&
\frac{x}{2R}\left[\frac{\bar{\gamma}_{\infty}}
{1+\bar{\gamma}_{\infty}}(\bar{\gamma}^{\,\prime}-\bp^{\,\prime})
-\frac{\bp^{\,\prime}\bp\,(2+\bp)}{(1+\bp)^2}\right]\nonumber\\
&+&\frac{x}{R^2}
\left(\bar{\gamma}-\frac{\bp^2}{1+\bp}\right),\end{eqnarray} with
a similar expression for $\bar{h}_{y\nu,}^{\;\;\;\;\;\nu}$
replacing $x$ with $y$ in the last equation. Here, $\bar{\gamma}$
is determined in terms of $\bp$ and its momentum by Eq.
(\ref{baga}). In particular, this value of $\bar{\gamma}$ is at
least quadratic in the fields of the reduced model.

It is easy to check that, while the metric derivatives
$\bar{h}_{\mu\nu,}^{\;\;\;\;\;\nu}$ do not generally vanish
according to Eq. (\ref{deDopr}), their value is in fact equal to
zero at linear order after the reduction of the system. As a
consequence, the gauge that we have chosen for the analysis of the
ER waves in linearized gravity is a de Donder gauge, and the gauge
fixing selected by Ashtekar and Pierri is a valid generalization
of it to the full (i.e. non-linearized) model. In fact, a
straightforward computation using the reduced metric (\ref{mink})
of the linearized model leads to the conclusion that
$\bar{h}_{\mu\nu}=2\bp\delta_{\mu}^{Z}\delta_{\nu}^{Z}$. Since
cylindrical symmetry guarantees the independence of the field
$\bp$ on the $Z$-coordinate, we see that our gauge for the
linearized theory satisfies the de Donder conditions, in agreement
with our comments above.

Actually, the relation between the linearization of the AP gauge,
on the one hand, and the de Donder gauge for the linearized ER
model, on the other, is even tighter. In general relativity, the
de Donder conditions select not just one, but a family of gauges,
leaving a remaining freedom in the choice of coordinates
\cite{MTW}. For ER waves in linearized gravity, however, it is
possible to see that the de Donder gauge is unique if one fixes
the location of the symmetry axis at $R=0$ and imposes there (the
linearized counterpart of) the regularity conditions (\ref{regu})
and (\ref{regu2}).

Let us prove this assertion. From Eq. (\ref{lic}) and the comments
above it, we see that, at linear order in the fields, the
regularity conditions (\ref{regu}) amount to
\begin{equation}\label{liregu}
\bar{\rho}(R\!=\!0)\!=\!\bar{\rho}^{\,\prime}(R\!=\!0)\!=
\!\bar{\gamma}(R\!=\!0)
\!=\! \bar{N}^R(R\!=\!0)\!=\!0.
\end{equation}
Besides, the stability under diffeomorphisms of the regularity
conditions demands the restriction (\ref{regu2}), which translates
into
\begin{equation}\label{liregu2}
\big(\bar{N}^{R}\big)^{\prime}\!(R\!= \!0)\!=0.\end{equation} On
the other hand, the de Donder conditions
$\bar{h}_{\mu\nu,}^{\;\;\;\;\;\nu}=0$ for the cylindrically
symmetric metric (\ref{neme}) are equivalent to the equations
\begin{eqnarray}\label{deDoeq1}
\hspace*{-.4cm}&&\big(\bar{N}^R\big)^{\prime}+\dot{\bar{N}\,}
-\frac{\dot{\bar{\gamma}\,}}{2}-
\frac{8G\dot{\bar{\rho}\,}}{R}=0,\\ \hspace*{-.4cm}&&
\!\dot{\,\,\bar{N}^R}+\bar{N}^{\prime}-
\frac{\bar{\gamma}^{\,\prime}}{2}-
\frac{\bar{\gamma}}{R}+
\frac{8G\bar{\rho}^{\,\prime}}{R}+\frac{8G\bar{\rho}}{R^2}=0.
\label{deDoeq2}\end{eqnarray}

The linearized Hamiltonian constraint in (\ref{licon}) and the
regularity conditions (\ref{liregu}) imply that
\begin{equation}\label{garo}
\bar{\gamma}=16G\bar{\rho}^{\,\prime}.\end{equation} Recalling the
Hamilton equations (\ref{lihameq}), the radial constraint of the
linearized theory is then straightforwardly satisfied. In
addition, differentiating Eqs. (\ref{deDoeq1}) and (\ref{deDoeq2})
with respect to $R$ and $t$, respectively, subtracting the
results, and substituting relation (\ref{garo}), we arrive at
\[
(\bar{N}^R)^{\prime\prime}-\!\ddot{\,\,\bar{N}^R}=0.\] So,
$\bar{N}^R$ can be expanded in terms of ``plane'' waves
$e^{ik(t+R)}$ and $e^{ik(t-R)}$, with $k\in\mathbb{R}$. But the
only superposition of these waves that satisfies the conditions
$\bar{N}^R=(\bar{N}^R)^{\prime}=0$ at $R=0$ for all times [see
requirements (\ref{liregu}) and (\ref{liregu2})] is the zero
field. For $\bar{N}^R=0$ and
$\bar{\gamma}=16G\bar{\rho}^{\,\prime}$, the de Donder equations
(\ref{deDoeq1}) and (\ref{deDoeq2}) reduce simply to
\begin{equation}\label{Neq}
\bar{N}=8G\bar{\rho}^{\,\prime}+\frac{8G\bar{\rho}}{R}+c,
\end{equation}
with $c$ being a constant.

We have not used yet the equation of motion for $\bar{\rho}$ in
the linearized system. This equation can be easily deduced from
the linearized action (\ref{liact}), taking into account the
Hamilton equations (\ref{lihameq}) and the form of the linearized
constraints:
\[
8G\ddot{\bar{\rho}\,}-\bar{N}^{\prime}-\dot{\;\bar{N}^R}=0.\] It
is possible to see that, given relation (\ref{garo}), the only
other independent equation of motion left in the system is that
for the field $\bp$, which percolates to the reduced model. With
$\bar{N}^R=0$ and formula (\ref{Neq}), the above equation for
$\bar{\rho}$ translates into
\[
\ddot{\bar{\rho}\,}=\bar{\rho}^{\prime\prime}+
\frac{\bar{\rho}^{\prime}}{R}-
\frac{\bar{\rho}}{R^2},\] which admits solutions of the form
$e^{ikt}J_1(Rk)$ and $e^{ikt}\,Y_{1}(Rk)$, $k\in\mathbb{R}$, in
terms of first-order Bessel functions \cite{AG}. Again, the only
solution of this type allowed by our regularity conditions at the
axis, which demand that $\bar{\rho}$ and $\bar{\rho}^{\,\prime}$
vanish at $R=0$, is the zero function. Since $\bar{\rho}$
vanishes, Eq. (\ref{Neq}) implies then that $\bar{N}$ must be
constant. This constant can be set equal to zero by requiring a
vanishing excess of the lapse with respect to the Minkowski
background, either at the symmetry axis or at spatial infinity.

In conclusion, we have shown that the de Donder conditions,
together with our regularity requirements at $R=0$, completely
determine the gauge choice for ER waves in linearized gravity. The
gauge fixing is such that all metric fields vanish except $\bp$.
The resulting reduced metric is just that found in Sec. IV for the
linearized model, namely the metric obtained with the
linearization of the gauge conditions selected by Ashtekar and
Pierri. As a consequence, the gauge chosen in the AP formulation
is a valid generalization of the de Donder gauge from the
linearized to the full ER model.

\section{Annihilation and creation variables}
\setcounter{equation}{0}

In order to discuss the connection between the Fock space employed
in the AP quantization of the ER model and that which would arise
in its perturbative quantization, we will analyze in this section
the relation between the annihilation and creation variables that
are associated with each of these two approaches.

As we have commented, the basic metric fields in standard
perturbative treatments of gravity, $\bar{h}_{\mu\nu}$, are linear
in the excess of the metric around Minkowski. In the linearized
theory, one chooses a gauge that simplifies the corresponding
equations of motion, for instance a de Donder gauge. When the
linearized description is modified by allowing the presence of
gravitational interactions, this gauge can be corrected with terms
that are of higher order in the fields, both to ensure that the
gauge continues to be well posed and to facilitate the analysis of
the system. In Sec. V we showed that the gauge choice made by
Ashtekar and Pierri is precisely a modified de Donder gauge of
this type. We will therefore select it as a valid gauge to compare
the results of the AP and the perturbative approaches to the
quantization of the ER model.

From this perspective, the relation between the two approaches is
based just on a field redefinition, namely the transformation
$\psi=\ln{(1+\bp)}$ mapping the field $\psi$ of the AP formulation
to the field $\bp$, which describes the difference with respect to
Minkowski of the diagonal $Z$-component of the metric. In the
reduced ER model, this field redefinition becomes a canonical
transformation when completed with the momentum change
(\ref{motra}). The relation that we are interested in is that
between the annihilation and creation variables associated with
each of the canonical pairs $(\psi,p_{\psi})$ and
$(\bp,\bar{P}_{\bp})$.

Although the fields considered are time dependent, we will see
below that, for our purposes, it will suffice to study their
relation on the initial time surface. If we wanted to analyze the
dynamics using the standard techniques of perturbative gravity, we
would be forced to consider the evolution in the auxiliary time
$T$ of the ER model, rather than in the physical time $t$, at
least in a first step. The reason is that the Lagrangian of the
reduced model is local only in the former case. For the physical
time, the action can be regarded as the sum of a local, a bilocal,
and in general multi-local terms of all orders. Once we had dealt
with the gravitational interactions in the auxiliary time by
perturbative methods, we could change the dynamical description to
the physical time in a second step, taking into account the
backreaction produced by the presence of gravitational waves in
the form of a deficit angle at spatial infinity, accompanied by a
modification of the norm of the asymptotic timelike Killing
vector.

This philosophy is in fact similar to that adopted in the
discussion of cylindrical gravitational waves with general
polarization as a Sigma model \cite{Ni,BMV2} (proposed as an
alternative approach to other quantization schemes \cite{KS}). In
this case, the gravitational action has also been made local with
the choice of an auxiliary time whose norm at spatial infinity,
though constant in the evolution, differs from the unity. The
change to the physical time is energy dependent, and leads to a
multi-local action.

It is instructive to see the expression of the reduced action of
the ER model corresponding to the auxiliary time $T$ in terms of
the two types of fields employed to describe the system, namely
$\psi$ and $\bp$. Remembering that, in the time $T$, the dynamical
generator is the Hamiltonian $H_0=\gamma_{\infty}/(8G_3)$ and
using relation (\ref{gamma}), one can check that the associated
reduced Lagrangian is
\begin{eqnarray}
L_0&=&\frac{1}{8G_3}
\!\int_0^{\infty}\!\!dR\,\frac{R}{2}\left[-(\psi^{\prime})^2+
(\partial_T\psi)^2\right]\nonumber\\
&=&\frac{1}{8G_3}\!\int_0^{\infty}\!\!dR\,\frac{R}{2(1+\bp)^2}
\left[-(\bp^{\prime})^2+
(\partial_T\bp)^2\right]\!.\nonumber\end{eqnarray} Thus, while the
AP formulation consists of a free-field parametrization of the
reduced system, described by $\psi$, the other parametrization,
natural from the viewpoint of a perturbative approach, leads to a
field $\bp$ with self-interactions of all orders, namely
\[
L_0\!=\!\sum_{n=0}^{\infty}(-1)^n\frac{(n+1)}{8G_3}\!\int_0^{\infty}
\!\!dR\, \frac{R}{2}\bp^n
\left[-(\bp^{\prime})^2+(\partial_T\bp)^2\right].\] In this sense,
one can interpret the AP formulation as a free-field realization
of the ER model.

Regardless of the time parameter selected to describe the
evolution of the ER model (the auxiliary or the physical one), the
initial time section of the system coincides in both cases, since
the times differ only by a positive normalization factor. When
quantizing the system, a Fock space is assigned to this initial
section. Being this space the same for the two natural choices of
Hamiltonian, both lead to unitarily equivalent Fock quantizations
\cite{BMV}. Therefore, in order to study the relation between the
Fock spaces of the AP and the perturbative approaches, we can
restrict all considerations just to the initial time surface
$T=t=0$, as we anticipated. Thus, from now on, by $\psi(R)$,
$p_{\psi}(R)$, $\bp(R)$, and $\bpp(R)$ we will understand the
initial values of these fields.

In the AP description, one introduces annihilation and creation
variables, $A(k)$ and $A^{\dagger}(k)$ ($k\in\mathbb{R}^+$),
corresponding to the expansion of the cylindrically symmetric
field $\psi$ in terms of zeroth-order Bessel functions. Employing
the form of $\psi$ on classical solutions given in Sec. II, the
Hamiltonian equation $\partial_T\psi=8Gp_{\psi}/R$ and the
identity
\begin{equation}\label{Besid}
\int_0^{\infty} \!\!dR\,Rk\,
J_0(Rk)J_0(R\tilde{k})=\delta(k,\tilde{k}),
\end{equation}
one can check that
\begin{equation}\label{annihi}
A(k)\!=\!\!\int_0^{\infty}\!\!\!dR\,\frac{J_0(Rk)}{2\sqrt{4G}}
\left[Rk\psi(R)+\!i8Gp_{\psi}(R)\right].\end{equation} The complex
conjugate of this relation provides $A^{\dagger}(k)$.

In fact, recalling that $\psi$ and $p_\psi$ are a canonical pair
of cylindrically symmetric fields, it is not difficult to see just
from Eq. (\ref{annihi}) that the only non-vanishing Poisson
brackets of $A(k)$ and $A^{\dagger}(k)$ are really
\[\{A(k),A^{\dagger}(\tilde{k})\}=-i\delta(k,\tilde{k}).
\]
Therefore, without appealing to the explicit form of the classical
solutions, we can regard formula $(\ref{annihi})$ and its complex
conjugate as the definition of a set of annihilation and creation
variables corresponding to the field $\psi$. Furthermore, the same
arguments apply exactly as well to any other cylindrically
symmetric field and its momentum as far as they form a true
canonical pair. For instance, we can adopt the point of view of
the perturbative approach and consider the pair $(\bp,\bpp)$ as
the fundamental canonical fields. Associated with them, we then
introduce the following type of annihilation variables:
\begin{equation}\label{annihi2}
a(k)\!=\!\int_0^{\infty}\!\!\!dR\,\frac{J_0(Rk)}{2\sqrt{4G}}
\left[Rk\bp(R)+\!i8G\bpp(R)\right],\end{equation} with their
complex conjugates providing the creation variables
$a^{\dagger}(k)$.

A point that is worth remarking is that the above definitions are
the natural ones from the perspective of the perturbative
approach. In the linearized gravitational theory, $\bp$ satisfies
the same cylindrical wave equation as $\psi$ does in the full
reduced ER model. Thus, the associated expansion of $\bp(R,T)$ in
terms of complex exponentials of $T$ and Bessel functions of $R$
leads precisely to the above annihilation and creation variables
in linearized gravity. In other words, the underlying mode
decomposition of the fields $\psi$ and $\bp$ is the same at first
perturbative order. Actually, since $(\bp,\bpp)$ and
$(\psi,p_{\psi})$ coincide at linear order, the series expansion
of $A(k)$ in powers of the over-barred fields leads from Eq.
(\ref{annihi}) to definition (\ref{annihi2}) as the leading term
in a perturbative expansion (the same line of reasoning applies to
the complex conjugate formulas for $A^{\dagger}$ and
$a^{\dagger}$).

Employing Eq. (\ref{Besid}), the introduced definitions of
creation and annihilation variables can be inverted to obtain the
initial values of the fields:
\begin{eqnarray}\label{fiac}
\!\!\!\!\psi(R)\!&=&\!\sqrt{4G}\!\int_0^{\infty}\!\!\!dk
J_0(Rk)[A(k)+A^{\dagger}(k)],\nonumber\\
\!\!\!\!p_{\psi}(R)\!&=&\!\!\frac{iR}{\sqrt{16G}}\!\int_0^{\infty}
\!\!\!dk\,k J_0(Rk)[-A(k)\!+\!A^{\dagger}(k)],\end{eqnarray} and
similar expressions for $\bp$ and $\bpp$. Note again that, in
arriving at these formulas, we have not used the explicit form of
the classical solutions. They are simply Bessel expansions of the
initial fields. The dynamics is encoded in the obviated evolution
of the annihilation and creation variables, which have been
restricted in our analysis to the initial time surface.

By combining Eqs. (\ref{annihi2}), (\ref{neps}), (\ref{motra}),
and (\ref{fiac}), it is now straightforward to deduce the highly
non-linear relation that exists between the particle-like
variables of the AP and the perturbative approaches:
\begin{eqnarray}\label{annianni}
\hspace*{-.4cm}&&a(k)\!=
\!\!\int_0^{\infty}\!\!\!\!dR\,\frac{J_0(Rk)}{2\sqrt{4G}}
\left[Rk\,\bp(R|A,\!A^{\dagger})\!+\!i8G\bpp(R|A,\!A^{\dagger})
\right],\nonumber\\
\hspace*{-.4cm}&&\bp(R|A,A^{\dagger})=e^{\psi(R|A,A^{\dagger})}
-1\nonumber\\
\hspace*{-.4cm}&&\hspace*{.8cm}:=\exp{\!\left\{\!\sqrt{4G}
\!\int_0^{\infty}
\!\!\!\!d\tilde{k}J_0(R\tilde{k})[A(\tilde{k})\!+\!
A^{\dagger}(\tilde{k})]\!\right\}}\!-\!1,\nonumber\\
\hspace*{-.4cm}&&i8G\bpp(R|A)=
i8Gp_{\psi}(R|A,A^{\dagger})e^{-\psi(R|A,A^{\dagger})}\nonumber\\
\hspace*{-.4cm}&&\hspace*{.8cm}:=\sqrt{4G}R\int_0^{\infty}\!\!\!
d\tilde{k}\,\tilde{k}J_0(R\tilde{k})[A(\tilde{k})-A^{\dagger}
(\tilde{k})]\nonumber\\ \hspace*{-.4cm}&&\hspace*{.8cm}\times\,
\exp{\left\{\!-\sqrt{4G}\!\int_0^{\infty}
\!\!\!d\breve{k}J_0(R\breve{k})[A(\breve{k})\!+\!
A^{\dagger}(\breve{k})]\right\}},
\end{eqnarray}
while $a^{\dagger}(k)$ is the complex conjugate of $a(k)$. Note
that in fact these definitions implement the reality condition
$\bp(R|A,A^{\dagger})>-1$ (ensuring that the induced metric is
positive definite) if $A(k)$ and $A^{\dagger}(k)$ are complex
conjugate to each other, because then $\psi(R|A,A^{\dagger})$ is
real everywhere. On the other hand, the inverse transformation
between the two sets of particle-like variables can be obtained by
substituting the relations $\psi=\ln{(1+\bp)}$ and
$p_{\psi}=\bpp(1+\bp)$ in formula (\ref{annihi}) and expressing
the initial values of $\bp$ and $\bpp$ in terms of $a(k)$ and
$a^{\dagger}(k)$ [using the analog of Eq. (\ref{fiac})]. Finally
we point out that, in formula (\ref{annianni}), each of the AP
annihilation and creation variables appears multiplied by a factor
of $\sqrt{G}$. As a result, one can understand the expansion of
$a(k)$ and $a^{\dagger}(k)$ in powers of such variables as
equivalent to a perturbative expansion in powers of $\sqrt{G}$.

\section{Regularized operators}
\setcounter{equation}{0}

Once one has established the relation between the sets of
annihilation and creation variables associated with the
perturbative analysis of the system and with the AP formulation, a
natural way to elucidate whether the two schemes lead to
equivalent Fock quantizations is the following. One can first try
to implement the variables $a(k)$ and $a^{\dagger}(k)$ (associated
with the perturbative approach) as annihilation and creation
operators acting on the Fock space of the AP quantization. The
perturbative vacuum would then be the (unique) state annihilated
by all the operators $\hat{a}(k)$. If this state is physical, i.e.
if its norm is finite, it determines the Fock space of the
perturbative approach. The two considered Fock quantizations would
then be unitarily equivalent, the equivalence being given by the
map from the perturbative to the AP vacuum. On the contrary, the
Fock quantizations would be inequivalent if the perturbative
vacuum is not normalizable.

Remembering relation (\ref{annianni}) (and its complex conjugate),
we might naively attempt to promote $a(k)$ and $a^{\dagger}(k)$ to
operators in the AP quantization by replacing the variables $A(k)$
and $A^{\dagger}(k)$ with their operator counterpart. However,
this procedure fails because, in the quantum version of expression
(\ref{annianni}), the fields $\hat{\psi}(R)$ and
$\hat{p}_{\psi}(R)$ that one obtains are not proper operators, but
operator-valued distributions \cite{AMM}. In particular, the
exponential of $\pm\hat{\psi}(R)$ is not rigorously defined.

These problems can be overcome by regularizing the fields. We will
only consider regularizations that consist of a cut-off $k_c$ in
wave-numbers (or equivalently, in momentum space), so that they
can be described by a regulator $g(k)$ of the form (\ref{cut}).
Recall that in this case $g^{2}(k)=g(k)$. The corresponding
regularized quantum field $\hat{\psi}(R|g)$ is given in Eq.
(\ref{regfi}) and is self-adjoint for every $k_c<\infty$. The
spectral theorem allows then to define the exponential
$e^{\pm\hat{\psi}(R|g)}$ as a positive operator \cite{AMM}.
Employing the Campbell-Baker Hausdorff formula
$e^{\hat{b}+\hat{c}}=e^{-[\hat{b},\hat{c}]/2}e^{\hat{b}}e^{\hat{c}}$,
which is valid for operators $\hat{b}$ and $\hat{c}$ whose
commutator is a $c$-number, one can see
\begin{equation}\label{noror}
e^{\pm\hat{\psi}(R|g)}=e^{2G||J_0(R\ast)g||}:e^{\pm\hat{\psi}(R|g)}:
\end{equation}
where the colon denotes normal ordering and
\[
||J_0(R\ast)g||=\int_0^{\infty} \!dk\, |J_0(Rk)g(k)|^2.\]

For the regularization of the product $i8Gp_{\psi}e^{-\psi}$ that
appears in Eq. (\ref{annianni}), we choose the ordering
\begin{eqnarray}\label{regpex}
\hspace*{-1.cm}&&i8G\left(\hat{p}_{\psi}(R|g)
e^{-\hat{\psi}(R|g)}\right)_{\!{\cal
N}} \!:=\!-\sqrt{4G}\,R\int_0^{k_c}dk\,k\,J_0(Rk)
\nonumber\\\hspace*{-1.cm}&&\;\;\;\times
\left[\hat{A}^{\dagger}(k)e^{-\hat{\psi}(R|g)}
-e^{-\hat{\psi}(R|g)}\hat{A}(k)\right].\end{eqnarray} Finally, we
define the following annihilation and creation-like operators
corresponding to $a(k)$ and $a^{\dagger}(k)$ for $k\leq k_c$:
\begin{eqnarray}\label{anniop}
\hspace*{-1.1cm}&&
\hat{a}(k|g):=\!g(k)\!\int_0^{\infty}\!\!\!\!dR\,\frac{J_0(Rk)}
{2\sqrt{4G}}
\nonumber\\ \hspace*{-1.1cm}&&\;\;\times\!
\left[Rke^{\hat{\psi}(R|g)}\!-\!R_g(G)k\!+\!i8G
\left(\hat{p}_{\psi}(R|g)
e^{-\hat{\psi}(R|g)}\right)_{\!{\cal N}}\right]\!,\nonumber\\
\hspace*{-1.1cm}&&
\hat{a}^{\dagger}(k|g):=\!g(k)\!\int_0^{\infty}\!\!\!\!dR\,
\frac{J_0(Rk)}{2\sqrt{4G}}
\nonumber\\ \hspace*{-1.1cm}&&\;\;\times\!
\left[Rke^{\hat{\psi}(R|g)}\!-\!R_g(G)k\!-\!i8G
\left(\hat{p}_{\psi}(R|g)
e^{-\hat{\psi}(R|g)}\right)_{\!{\cal N}}\right]\!.
\end{eqnarray}
By construction, these operators are adjoint to each other. In
order to account for part of the order ambiguity, we have left the
freedom to represent $R$ quantum mechanically by a $c$-number
$R_g(G)$ that may depend on the cut-off, as well as on the quantum
gravitational constant $G$. To recover the semiclassical limit, we
impose the condition that $R_g$ tends to $R$ in the limit
$G\rightarrow 0$. For simplicity, we also assume that $R_g(G)$ is
analytic in $G$.

The commutators of the operators (\ref{anniop}) are computed in
the Appendix. Of course, they do not depend on the form of the
$c$-number $R_g(G)$. We will only comment two important properties
of these commutators. First, using that $e^{\pm\hat{\psi}(R|g)}$
tends to the identity operator in the limit of vanishing $G$ and
remembering the integral expression (\ref{Besid}), one can check
that the only non-vanishing commutators of our operators when
$G\rightarrow 0$ are
\[
\lim_{G\rightarrow 0}
\left[\hat{a}(k|g),\hat{a}^{\dagger}(\tilde{k}|g)\right]
=g(k)\delta(k,\tilde{k}).\] So, in this kind of semiclassical
limit, we recover the algebra of a set of annihilation and
creation operators in the region of wave-numbers to which we are
restricting our analysis. Second, one can proceed to remove the
cut-off by taking the limit $g(k)\rightarrow 1$ or, equivalently,
$k_c\rightarrow\infty$. Assuming that this limit can be taken
inside the integrals in expression (\ref{commu}) and using the
identity (\ref{Besid}), a careful calculation shows that
\begin{eqnarray}
\lim_{g\rightarrow
1}\,\left[\hat{a}(k|g),\hat{a}(\tilde{k}|g)\right]&=&
\lim_{g\rightarrow 1}
\left[\hat{a}^{\dagger}(k|g),\hat{a}^{\dagger}(\tilde{k}|g)\right]
=0,\nonumber\\ \lim_{g\rightarrow 1}
\left[\hat{a}(k|g),\hat{a}^{\dagger}(\tilde{k}|g)\right]&=&
\delta(k,\tilde{k}).
\nonumber\end{eqnarray} In this sense, the desired algebra of
annihilation and creation operators associated with the
perturbative quantization scheme can be regarded as the limit of
our algebra of operators when the cut-off is driven to infinity.

Substituting the regulated expressions (\ref{regfi}) and
(\ref{regpex}) in the definition of the operators $\hat{a}(k|g)$
and $\hat{a}^{\dagger}(k|g)$, it is not difficult to obtain their
expansion in powers of annihilation and creation operators of the
AP quantization, $\hat{A}(k)$ and $\hat{A}^{\dagger}(k)$.
According to our comments at the end of Sec. VI, this expansion
reproduces the power series in $\sqrt{G}$, except for the possible
$G$ dependence introduced by the $c$-number $R_g(G)$, which
(partially) accounts for operator ordering ambiguities. Therefore,
one can interpret the series in $\sqrt{G}$ in the sense that each
additional power corresponds (in a certain operator ordering) to
the creation or annihilation of an extra particle in the AP
quantization. Explicitly, the series will have the form
\begin{eqnarray}\label{gexp}
\hat{a}(k|g)&=&\sum_{n=0}^{\infty}
(G)^{n/2}\hat{a}_{(n)}(k|g),\nonumber\\
\hat{a}^{\dagger}(k|g)&=&\sum_{n=0}^{\infty}
(G)^{n/2}\hat{a}^{\dagger}_{(n)}(k|g),\end{eqnarray} where now the
operators $\hat{a}_{(n)}(k|g)$ and their adjoints are independent
of the quantum gravitational constant $G$.

Remembering that $R_g(G)$ is analytic in $G$ and equal to $R$ at
$G=0$, and using identity (\ref{Besid}), it is easy to find the
zeroth-order contribution to $\hat{a}(k|g)$:
\begin{eqnarray}\label{azero}
\hat{a}_{(0)}(k|g)\!\!&=&\!
\!g(k)\!\int_0^{\infty}\!\!\!\!dR\,\frac{J_0(Rk)}{2\sqrt{4G}}
\left[Rk\hat{\psi}(R|g)\!+\!i8G
\hat{p}_{\psi}(R|g)\right]\nonumber\\
&=&\!g(k)\hat{A}(k).\nonumber
\end{eqnarray}
Similarly,
$\hat{a}^{\dagger}_{(0)}(k|g)=g(k)\hat{A}^{\dagger}(k)$. Thus, our
definition of annihilation and creation-like operators for the
perturbative approach is such that, in the considered sector of
wave-numbers $k\leq k_c$, they coincide with the annihilation and
creation operators of the AP formulation at dominant order in
$\sqrt{G}$. In other words, at first perturbative order the two
types of particles can be identified in the region of momentum
space below the cut-off. Note that one can completely determine
the particle contain in this perturbative limit by finally
proceeding to remove the regulator, i.e.
$\hat{A}(k)=\lim_{g\rightarrow 1}\lim_{G\rightarrow
0}\hat{a}(k|g)$ for all $k\in \mathbb{R}$.

\section{The perturbative vacuum}
\setcounter{equation}{0}

We have seen that the algebra of $\hat{a}(k|g)$ and
$\hat{a}^{\dagger}(k|g)$, in the limit of infinite cut-off, has
the form of that of a set of annihilation and creation operators.
In addition, their values for $G=0$ coincide with the annihilation
and creation operators of the AP formulation for wave-numbers $k$
smaller than the cut-off, reproducing the whole set of those
operators when the regulator disappears. The AP vacuum $|0\rangle$
is hence totally fixed by the condition that it be annihilated by
the operators $\hat{a}_{(0)}(k|g)$ for every value of the cut-off
$k_c$, or equivalently in the limit $k_c\rightarrow\infty$.
Besides, it is possible to check that $|0\rangle$ is not
annihilated by all the operators $\hat{a}(k|g)$ for each fixed
$k_c>0$, and so neither is it when the regulator is removed.
Therefore, the AP vacuum differs from that corresponding to the
perturbative approach, which we will call $|\bar{0}\rangle$.

Provided that the latter of these vacua belongs to the Fock space
of the AP quantization, a way to determine it is the following.
First, for each fixed cut-off, we find a state $|\bar{0}_g\rangle$
annihilated by all the operators $\hat{a}(k|g)$ and such that
coincides with the AP vacuum $|0\rangle$ in the limit
$G\rightarrow 0$. If $|\bar{0}_g\rangle$ is a physical state, one
can choose it with unit norm. The vacuum $|\bar{0}\rangle$ would
then be attained as the limit of this normalized state when the
regulator is removed, $k_c\rightarrow\infty$.

Note however that our definition (\ref{anniop}) of $\hat{a}(k|g)$
involves only annihilation and creation operators of the AP
formulation with $k\leq k_c$. As a consequence, one can anticipate
the existence of ambiguities in the determination of
$|\bar{0}_g\rangle$, owing to a lack of uniqueness in the allowed
contributions from the sector of AP particles with wave-number
greater than the cut-off. Nonetheless, this ambiguity can be
eliminated by demanding that the regularized perturbative vacuum
$|\bar{0}_g\rangle$ has no projection in that sector. This is a
natural condition if we interpret the regularization as the
removal of all interactions and particles with energies above the
cut-off. In particular, it is consistent with the requirement that
the limit of $|\bar{0}_g\rangle$ when $G$ tends to zero be the AP
vacuum $|0\rangle$, because this is the only physical state which
does not contain particles with $k>k_c$ and is annihilated by all
the operators $\hat{a}_{(0)}(k|g)$ [which are equal to
$\hat{A}(k)$ below the cut-off and vanish otherwise].

The computation of $|\bar{0}_g\rangle$ can be carried out
perturbatively in terms of the quantum gravitational constant $G$.
In order to do it, one employs the series (\ref{gexp}) for the
annihilation-like operators $\hat{a}(k|g)$ and expands the
regularized vacuum as a formal power series of $\sqrt{G}$ as well:
\begin{equation}\label{vexp}
|\bar{0}_g\rangle=|0\rangle +\sum_{n=1}^{\infty}(G)^{n/2}
\,|\Phi_{n,g}\rangle.\end{equation} Here, we have made explicit
that the dominant contribution must be the vacuum $|0\rangle$, and
$|\Phi_{n,g}\rangle$ designates a linear superposition of states
with a finite, non-zero number of AP particles whose wave-number
is bounded by the cut-off, $k\leq k_c$. The above formula can
equivalently be regarded as providing the AP vacuum $|0\rangle$ as
a formal series in $\sqrt{G}$ in terms of the regularized
perturbative vacuum, $|0\rangle=|\bar{0}_g\rangle -\sum (G)^{n/2}
|\Phi_{n,g}\rangle$.

By an iterative process, one can deduce the form of all the states
$|\Phi_{n,g}\rangle$. Namely, once $|\Phi_{n,g}\rangle$ is known
for every $n<m$, one can determine $|\Phi_{m,g}\rangle$ from the
condition that $\hat{a}(k|g)|\bar{0}_g\rangle$ vanish for all $k$
at order $G^{m/2}$. Let us consider the case $m=1$,
\begin{equation}\label{zero}
\hat{a}_{(0)}(k|g)|\Phi_{1,g}\rangle+\hat{a}_{(1)}(k|g)|0\rangle=0.
\end{equation}
From Eq. (\ref{anniop}) and the analyticity of $R_g(G)$ in $G$,
one can see that
\begin{eqnarray*}
\hspace*{-.4cm}&&\hat{a}_{(1)}(k|g)|0\rangle=\frac{g(k)}{2}
\,\left[\int_0^{\infty}\!\!dR\,J_0(Rk)\,k\,
E_g(R)\,|0\rangle\right.\\ \hspace*{-.4cm}&&
+\!\left.\!\int_0^{k_c}\!\!\!dk_1\!\!
\int_0^{k_c}\!\!\!dk_2\left(k_1\!+\!k_2\!+\!k\right)\!F(k_1,k_2,k)
\hat{A}^{\dagger}(k_1)\hat{A}^{\dagger}(k_2)|0\rangle\right]\!,
\end{eqnarray*}
where we have defined the functions
\begin{eqnarray}
\hspace*{-1.cm}&&E_g(R):=||J_0(R\ast)g||\;R-\frac{1}{2}\partial_G
R_g(G\!=\!0),\nonumber\\ \label{fks}
\hspace*{-1.cm}&&F(k_1,k_2,k_3)\!:=\!\!\int_0^{\infty} \!\!\!dR
\,R\,J_0(Rk_1)J_0(Rk_2) J_0(Rk_3),\end{eqnarray} and we have
interchanged the order of integration in $R$ and in $(k_1,k_2)$.
The first term in $E_g(R)$ arises from the derivative with respect
to $G$ of the factor $e^{2G||J_0(R\ast)g||}$, that appears in the
operator $e^{\hat{\psi}(R|g)}$ when expressed in normal ordering
[see Eq. (\ref{noror})]. On the other hand, notice that the
function $F(k_1,k_2,k_3)$ is symmetric in all its arguments.

Condition (\ref{zero}) implies then that
\begin{eqnarray}
\hspace*{-.3cm}|\Phi_{1,g}\rangle\!&=&\!
-\int_0^{k_c}\!\!dk_1\,\frac{k_1}{2}
\int_0^{\infty}\!\!dRJ_0(Rk_1)\,
E_g(R)\hat{A}^{\dagger}(k_1)|0\rangle\nonumber\\
&-&\!|\Upsilon_g\rangle,\nonumber
\end{eqnarray}
with
\begin{eqnarray}
|\Upsilon_g\rangle\!&:=&\!
\int_0^{k_c}\!\!dk_1\!\int_0^{k_c}\!\!dk_2\!
\int_0^{k_c}\!\!dk_3\,\frac{\left(k_1\!+\!k_2\!+\!k_3\right)}{6}\,
\!F(k_1,k_2,k_3)\nonumber\\
&\times&\!\hat{A}^{\dagger}(k_1)\hat{A}^{\dagger}(k_2)
\hat{A}^{\dagger}(k_3)|0\rangle.\nonumber
\end{eqnarray}
Any possible contribution to $|\Phi_{1,g}\rangle$ proportional to
the vacuum has been obviated, because it is not necessary to
satisfy condition (\ref{zero}). Moreover, such a contribution can
always be absorbed in the dominant term of the series expansion
(\ref{vexp}) up to a $G$-dependent, global numeric factor in
$|\bar{0}_g\rangle$ that only changes the norm of this state.

Similar arguments can be applied to fix the next correction to the
AP vacuum, $|\Phi_{2,g}\rangle$, as a superposition of states with
a finite but non-zero number of particles belonging to the sector
$k\leq k_c$, using the condition
\[
\hat{a}_{(0)}(k|g)|\Phi_{2,g}\rangle+\hat{a}_{(1)}(k|g)|\Phi_{1,g}
\rangle+\hat{a}_{(2)}(k|g)|0\rangle=0.
\]
Likewise, $|\Phi_{m,g}\rangle$ can be fixed from the corresponding
condition at order $G^{m/2}$ once $\{|\Phi_{n,g}\rangle;\,n<m\}$
have been determined.

Employing that $|\Phi_{1,g}\rangle$ is the sum of an one-particle
state and the three-particles state $|\Upsilon_g\rangle$, which
are orthogonal to each other and to the AP vacuum, and the fact
that $|\Phi_{2,g}\rangle$ is a linear combination of states with
non-zero particles, so that $\langle 0|\Phi_{2,g}\rangle= 0$, we
conclude that, up to corrections of order $G^{n/2}$ with $n\geq
3$, the norm of $|\bar{0}_g\rangle$ satisfies
\[
\langle\bar{0}_g|\bar{0}_g\rangle \geq \langle 0|0\rangle +G
\langle\Upsilon_g|\Upsilon_g\rangle=1+G
\langle\Upsilon_g|\Upsilon_g\rangle.
\]

We will now show that the norm of $|\Upsilon_g\rangle$ is infinite
regardless of the value of the cut-off, and therefore
$|\bar{0}_g\rangle$ is not a physical state (at least as a power
series in $\sqrt{G}$). Thus, one cannot reach in this way a
normalized perturbative vacuum in the limit
$k_c\rightarrow\infty$. This strongly indicates that the
perturbative vacuum is not included in the Fock space of the AP
formulation, implying that the two discussed approaches to the
quantization of the ER waves are unitarily inequivalent. In any
case, the non-normalizability of $|\Upsilon_g\rangle$ means that
the perturbative vacuum is not analytic in $\sqrt{G}$,
invalidating the perturbative calculation presented above.

The norm of $|\Upsilon_g\rangle$ is given by
\[
\langle\Upsilon_g|\Upsilon_g\rangle\!=\!\!\!
\int_0^{k_c}\!\!\!\!dk_1\!\int_0^{k_c}\!\!\! \!dk_2\!
\int_0^{k_c}\!\!\!\!dk_3\,
\frac{\left(k_1\!+\!k_2\!+\!k_3\right)^2\!}{6}\,
\!F^2(k_1,\!k_2,\!k_3).\] So, in order to obtain it, we will first
calculate the integral (\ref{fks}) that provides the function
$F(k_1,k_2,k_3)$. This integral can be computed explicitly, e.g.,
using the formulas of Ref. \cite{GR}. The result is
\begin{eqnarray}
\hspace*{-.5cm}F(k_1,k_2,k_3)&\!=
\!&\Theta(k_1+k_2-k_3)\Theta(k_3-|k_1-k_2|)
\nonumber\\
\hspace*{-.5cm}&\!\times\!&\frac{2}{\pi\sqrt{4k_1^2k_2^2
-(k_1^2+k_2^2-k_3^2)^2}},\nonumber\end{eqnarray} where $\Theta(k)$
is the Heaviside step-function, equal to the unity if $k$ is
positive and vanishing otherwise. One therefore arrives at
\begin{eqnarray}
\hspace*{-.5cm}&&\langle\Upsilon_g|\Upsilon_g\rangle\!=
\!\!\int_0^{k_c}\!\!\!dk_1\!\int_0^{k_c}\!\!\!dk_2\!
\int_{|k_1-k_2|}^{{\rm min}\{k_c,k_1+k_2\}}\!\!dk_3\,
\frac{\left(k_1\!+\!k_2\!+\!k_3\right)}{6\pi^2} \nonumber\\
\hspace*{-.5cm}&&\hspace*{.5cm}
\times\frac{1}{(k_1+k_2-k_3)(k_3-|k_1-k_2|)(k_3+|k_1-k_2|)}.
\nonumber\end{eqnarray} Here, ${\rm min}\{a,b\}$ denotes the
minimum of the numbers $a$ and $b$.

Note that the integrand in the above expression is positive in the
integration region, and that the last integral has a simple pole
at the boundary $k_3=|k_1-k_2|$ of the integration interval for
$k_3$. As a consequence, the integral that determines the norm of
$|\Upsilon_g\rangle$ diverges for all positive values of the
cut-off $k_c$. In other words, regardless of the cut-off,
$|\Upsilon_g\rangle$ is not a physical state. We thus conclude
that the vacuum of the perturbative approach is not accesible as a
power series in $\sqrt{G}$ in the Fock space of the AP
quantization.

It is worth emphasizing that the divergence of the norm of
$|\Upsilon_g\rangle$ does not arise as a result of taking the
limit in which the cut-off is removed because, if that were the
case, one could proceed to renormalize the perturbative vacuum.
Namely, one could first normalize $|\bar{0}_g\rangle$, obtaining
the unit norm state $|\tilde{0}_g\rangle=|\bar{0}_g\rangle [
\langle\bar{0}_g|\bar{0}_g\rangle]^{-1/2} $, and only then
consider the limit $k_c\rightarrow\infty$.

\section{Summary and conclusions}
\setcounter{equation}{0}

We have investigated the relation between the notion of particle
that arises in the quantum framework developed by Ashtekar and
Pierri for the description of the ER waves and that expected in a
perturbative approach to the quantization of this system. We have
started by introducing a set of metric fields that are specially
suitable for the analysis of the model in linearized gravity,
since the fields are linear in the excess of the metric with
respect to Minkowski. Using this new field parametrization of the
ER spacetimes, we have discussed their quantization in the
linearized theory of gravity. We have shown that the linearization
of the gauge fixing conditions adopted by Ashtekar and Pierri
provides the only gauge choice of the de Donder type which
respects the regularity conditions imposed on the metric at the
axis of rotational symmetry, located at a fixed location [namely,
the origin of the radial coordinate $R$]. This result allows one
to interpret the gauge selected by Ashtekar and Pierri as a
well-posed generalization to the ER model of the de Donder gauge
compatible with the regularity at the axis.

From this perspective, the perturbative description of the system
can be made to rest on a field $\bp$, that parametrizes the metric
in the AP gauge and is linear in the excess around Minkowski. By
contrast, the parametrization chosen in the AP formulation is
based on a field $\psi$ that is highly non-linear in the metric
excess, but straightforwardly incorporates the reality conditions
on the metric and, more importantly, is subject to a linear
(reduced) dynamics. The correspondence between the two fields can
be completed into a canonical transformation on the phase space of
the reduced ER model. This canonical transformation provides the
key relation for discussing the correspondence between the
particle-like variables of the two considered descriptions.

Given a field and its canonical momentum, both possessing
rotational symmetry and being regular at the symmetry axis, it is
possible to expand their initial values in terms of zeroth-order
Bessel functions. Using this property, one can associate with the
canonical pair of axisymmetric fields on a constant time section a
set of annihilation and creation variables with positive
wave-numbers. This possibility is at hand both for the field
$\psi$ and its momentum in the AP formulation and for the field
$\bp$ and its momentum in the perturbatively inspired description
that is linear in the metric excess. The transformation between
both pairs of fields provides the relation between the
corresponding sets of annihilation and creation variables.
Furthermore, since the linearization of both parametrizations is
the same, the introduced particle-like variables coincide at
linear order.

What one gets in this way is the expression, e.g., of the
particle-like variables associated with the perturbative analysis
as highly non-linear functionals of the corresponding set of AP
variables. This expression can be expanded as a power series in
the latter set, each new particle-like variable being accompanied
by a factor of $\sqrt{G}$. Thus, one can regard the square root of
the quantum gravitational constant as the interaction constant of
the model, and the expansion in the number of created and
annihilated particles as a perturbative expansion in that
constant. As we have said, the expansion is such that the
annihilation and creation variables of the AP formulation are
taken as the leading contribution for small $\sqrt{G}$, i.e. the
mode decomposition is made to coincide with that of the AP
description at dominant perturbative order.

We have next proceeded to consider the quantum version of the
relation between the variables of annihilation and creation type
for the two formulations. We have employed as starting point the
Fock space of the AP quantization, since it provides a
mathematically well-posed framework where the quantum issues can
be discussed with rigor. The first problem that has been necessary
to overcome, in order to promote to meaningful operators the
particle-like variables of the perturbative scheme, is to
regularize the quantum fields. This has been done by introducing a
cut-off in the model and suppressing all particle interactions
with energies (or equivalently wave-numbers) above it. The
resulting operators have been proved to reproduce a formal algebra
of annihilation and creation operators in the limit in which the
cut-off is removed. In addition, the representation chosen is such
that, for vanishing quantum gravitational constant $G$, one
exactly recovers the annihilation and creation operators of the AP
quantization in the sector of particles with wave-numbers below
the cut-off.

Using these regularized operators, one can investigate whether the
perturbative vacuum can be represented as a physical state in the
Fock space of the AP quantization. If this were the case, the two
considered quantum theories (namely, those based on the AP and on
the perturbative vacuum) would be unitarily equivalent. In more
detail, we wanted to elucidate whether there exists a physical
state in the Fock space of the AP formalism that can play the role
of vacuum in the perturbative approach and be reached in the limit
of infinite cut-off as a perturbative power series in the coupling
constant $\sqrt{G}$, with interacting-free term (i.e., the
contribution at $G=0$) given by the original AP vacuum.

The other terms in this (regularized) series for the perturbative
vacuum can be determined, for each fixed value of the cut-off, by
imposing the condition that they consist of linear superpositions
of states with a finite but non-zero number of AP particles, and
that they be annihilated by all the regularized annihilation
operators. In particular, we have studied in detail the first
perturbative correction, proportional to $\sqrt{G}$. It is formed
by a three-particles and an one-particle states. More importantly,
we have proved that the norm of the three-particles state is
infinite regardless of the value of the cut-off. Therefore, no
normalizable perturbative vacuum is accesible from the AP vacuum
as a power series in $\sqrt{G}$.

This result is a clear indication of the inequivalence of the Fock
quantizations associated with the AP and the perturbative
approaches. Furthermore, the fact that the perturbative vacuum
cannot be realized as a physical state analytic in $\sqrt{G}$
prevents one from applying standard perturbative calculations
based on the number of AP particles involved in the interaction,
so that a naive perturbative treatment of the system is bound to
fail.

As we have commented, the Fock space that we have considered
describes the reduced degrees of freedom of the system in a
section of constant initial time. In this space one can introduce
a quantum dynamics, which provides the evolution of these degrees
of freedom as time progresses. In the case of the ER waves, this
notion of reduced dynamics can be linked to two types of quantum
Hamiltonian, a local one that describes the evolution in an
auxiliary time, conformally flat in two dimensions together with
the radial coordinate, and a non-local one that corresponds to a
physical time, normalized to the unity at spatial infinity. The
analysis of the system with the first Hamiltonian is trivial in
the AP quantization, in the sense that it leads to a free-field
realization. The quantum evolution in the physical time, on the
other hand, is much more involved. It can also be studied by a
perturbative approach, but this time the perturbative order
corresponds to the degree of non-locality. An $n$-point
contribution to the Hamiltonian will be proportional to the
$(n-1)$-th power of the gravitational constant $G$.

The fact that the same constant $G$ plays the role of perturbative
parameter in both types of analyses (one linked to the number of
AP particles involved in the local interaction and the other to
the degree of non-locality in the dynamics) may lead to some
confusion. This is in part due to the fact that the system
possesses only a fundamental constant, so that it will show up in
any natural expansion. Anyway, a full perturbative analysis can
always be made in two steps, first discussing the quantum system
at a fixed instant of time, as we have made here, and then taking
into account the non-locality introduced by the change from the
auxiliary to the physical time. This second type of perturbative
issues will be considered elsewhere \cite{BMV3}.

\appendix
\renewcommand{\theequation}{A.\arabic{equation}}

\section{}
\setcounter{equation}{0}

In this appendix we compute the commutators of the operators
$\hat{a}(k|g)$ and $\hat{a}^{\dagger}(k|g)$ defined in Eq.
(\ref{anniop}). Let us introduce the symbolic notation
$\hat{a}^{\ddag}(k|g)$ for both types of operators, with
$(-1)^{\ddag}$ equal to $-1$ and 1, respectively, in the
annihilation and creation case. Using relation (\ref{noror}) and
the basic commutators (\ref{ancre}), one can then see that
\begin{eqnarray}\label{commu}
\hspace*{-.5cm}&&\left[\hat{a}(k|g),\hat{a}^{\ddag}(\tilde{k}|g)\right]
\!\!=\!g(k)g(\tilde{k})\!\!\int_0^{\infty}\!\!\!\!dR\frac{R}{2}J_0(Rk)
\!\!\int_0^{\infty}\!\!\!\!d\tilde{R}\tilde{R}
J_0(\tilde{R}\tilde{k})\nonumber\\
\hspace*{-.5cm}&&\;\times\int_0^{k_c}\!\!dk_1\,k_1J_0(Rk_1)
\int_0^{k_c}\!\!dk_2\,J_0(\tilde{R}k_2)\left[\,(-1)^{\ddag}\,
\sqrt{4G}\,k_2\right. \nonumber\\
\hspace*{-.5cm}&&\;\times\!\left(\!\left\{\!J_0(Rk_2)
\hat{A}^{\dagger}(k_1)\!-\!J_0(\tilde{R}k_1)
\hat{A}^{\dagger}(k_2)\right\}\! e^{-\hat{\psi}(R|g)}
e^{-\hat{\psi}(\tilde{R}|g)}\right.\nonumber\\ \hspace*{-.5cm}&&\;
\left.+\, e^{-\hat{\psi}(R|g)}
e^{-\hat{\psi}(\tilde{R}|g)}\left\{J_0(\tilde{R}k_1)
\hat{A}(k_2)-J_0(Rk_2)\hat{A}(k_1)\right\}\right)\nonumber\\
\hspace*{-.5cm}&&\;+\!\left.\delta(k_1,k_2)\!
\left\{\tilde{k}e^{-\hat{\psi}(R|g)}
e^{\hat{\psi}(\tilde{R}|g)}\!+\!(-1)^{\ddag}ke^{\hat{\psi}(R|g)}
e^{-\hat{\psi}(\tilde{R}|g)}\!\right\}\!\right]\!.\nonumber\\
\hspace*{-.5cm}&&\;\end{eqnarray}

\begin{acknowledgments}
The authors are greatly thankful to M. Varadarajan for suggestions
and enlightening conversations. They are also grateful to A.
Ashtekar and A. Corichi for valuable questions and discussions.
This work was supported by the Spanish MCYT projects
BFM2002-04031-C02 and BFM2001-0213.
\end{acknowledgments}


\begin{thebibliography}{35}

\bibitem{Be} G. Beck, Z. Phys. {\bf 33}, 713 (1925).

\bibitem{ER} A. Einstein and N. Rosen, J. Franklin Inst. {\bf 223},
43 (1937).

\bibitem{WCS} K.S. Thorne, Phys. Rev. {\bf 138}, B251 (1965); M.A.
Melvin, Phys. Rev. {\bf 139}, B225 (1965).

\bibitem{Ch} See Appendix B in S. Chandrasekhar, Proc. R. Soc. Lond.
A {\bf 408}, 209 (1986).

\bibitem{EK} P. Jordan, J. Ehlers, and W. Kundt, Akad. Wiss.
Lit. Mainz Abh. Math. Naturwiss. Kl. {\bf 1960} No. 2 (1960); A.S.
Kompaneets, Zh. \'{E}ksp. Teor. Fiz. {\bf 34}, 953 (1958) [Sov. Phys.
JETP {\bf 7}, 659 (1958)].

\bibitem{Ku} K. Kucha\v{r}, Phys. Rev. D {\bf 4}, 955 (1971).

\bibitem{Al} M. Allen, Class. Quantum Grav. {\bf 4}, 149
(1987).

\bibitem{AP} A. Ashtekar and M. Pierri, J. Math. Phys. {\bf 37}, 6250
(1996).

\bibitem{Ma} M. Varadarajan, Class. Quantum Grav. {\bf 17},
189 (2000).

\bibitem{AMM} M.E. Angulo and G.A. Mena Marug\'{a}n, Int. J. Mod. Phys. D
{\bf 9}, 669 (2000).

\bibitem{Ash} A. Ashtekar, Phys. Rev. Lett. {\bf 77}, 4864 (1996).

\bibitem{GP} R. Gambini and J. Pullin, Mod. Phys. Lett. A
{\bf 12}, 2407 (1997).

\bibitem{DT} A.E. Dom\'{\i}nguez and M.H. Tiglio, Phys. Rev. D
{\bf 60}, 064001 (1999).

\bibitem{BMV} J.F. Barbero G., G.A. Mena Marug\'{a}n, and E.J.S.
Vi\-llase\~{n}or, Phys. Rev. D {\bf 67}, 124006 (2003).

\bibitem{Ni2} M. Niedermaier, Phys. Lett. B {\bf 498}, 83 (2001).

\bibitem{KHB} I. Kouletsis, P. H\'{a}j\'{\i}\v{c}ek, and J. Bi\v{c}\'{a}k, Phys.
Rev. D {\bf 68}, 104013 (2003).

\bibitem{RT} J.D. Romano and C.G. Torre, Phys. Rev. D {\bf 53}, 5634
(1996).

\bibitem{AV} A. Ashtekar and M. Varadarajan, Phys. Rev. D {\bf
50}, 4944 (1994).

\bibitem{Va} M. Varadarajan, Phys. Rev. D {\bf 52}, 2020
(1995).

\bibitem{ABS} A. Ashtekar, J. Bi\v{c}\'{a}k, and B.G. Schmidt,
Phys. Rev. D {\bf 55}, 669 (1997); {\bf 55}, 687 (1997).

\bibitem{Gu} G.A. Mena Marug\'{a}n, Phys. Rev. D {\bf 63}, 024005
(2001).

\bibitem{MMM} N. Manojlovi\'{c} and G.A. Mena Marug\'{a}n, Class.
Quantum Grav. {\bf 18}, 2065 (2001).

\bibitem{MTW} See, e.g., C.W. Misner, K.S. Thorne, and J.A. Wheeler,
{\it Gravitation} (Freeman, San Francisco, 1973).

\bibitem{Di} Strictly speaking, we then have $(R/G)\in \mathbb{R}^+$,
$(Z/G)\in \mathbb{R}$, and $(t/G)\in \mathbb{R}$.

\bibitem{Ma2} We thank M. Varadarajan for remarking this point.

\bibitem{AG}
{\it Handbook of Mathematical Functions}, Natl. Bur. Stand. Appl.
Math. Ser. No. 55, revised 9th. ed., edited by M. Abramowitz and
I.A. Stegun (U.S. Govt. Print. Off., Washington D.C., 1970).

\bibitem{RS} M. Reed and B. Simon, {\it Methods of Modern
Mathematical Physics II: Fourier Analysis, Self Adjointness}
(Academic Press, New York, 1975).

\bibitem{Ni} M. Niedermaier, JHEP {\bf 12}, 066 (2002);
Nucl. Phys. B {\bf 673}, 131 (2003).

\bibitem{BMV2} J.F. Barbero G., G.A. Mena Marug\'{a}n, and E.J.S.
Villa\-se\~{n}or, Int. J. Mod. Phys. D (in press), gr-qc/0402096.

\bibitem{KS} D. Korotkin and H. Samtleben, Phys. Rev. Lett. {\bf
80}, 14 (1998).

\bibitem{GR} Employ formula 8 of chapter 6.57 and formula 8.704 in
I.S. Grashteyn and I.M. Ryzhik, {\it Table of Integrals, Series
and Products}, 5th. ed. (Academic Press, London, 1994). For the
form of the hypergeometric function involved in these expressions
one can use formula 15.1.12 of \cite{AG}.

\bibitem{BMV3} J.F. Barbero G., G.A. Mena Marug\'{a}n, and E.J.S.
Villa\-se\~{n}or, J. Math. Phys. (in press), gr-qc/0405075.

\end{thebibliography}
\end{document}